\documentclass[12pt]{iopart}
\usepackage{graphicx}
\usepackage{color}

\newcommand{\gtsimeq}{\raisebox{-0.6ex}{$\,\stackrel
        {\raisebox{-.2ex}{$\textstyle >$}}{\sim}\,$}}

\begin{document}
\title{Collisions of bright solitary matter waves}
\author{N.G. Parker$^1$\footnote{Current address: Department of Physics and Astronomy, McMaster University, Hamilton, Canada }, A. M. Martin$^1$, S. L. Cornish$^2$ and C. S. Adams$^2$}
\address{$^1$ School of Physics, University of Melbourne, Parkville,
Victoria 3010, Australia
\\$^2$ Department of Physics, Durham University, Durham, DH1 3LE, UK}
\ead{n.p@physics.org}
\begin{abstract}
The collisions of three-dimensional bright solitary matter waves formed from atomic
Bose-Einstein condensates are shown to exhibit rich behaviour.  Collisions range from being elastic to completely destructive due to the onset of collapse during the interaction.  Through a detailed quantitative analysis we map out the role of
relative phase, impact speed and interaction strength. In particular, we identify the importance of the collapse time in the onset of unstable
collisions and show how the relative phase controls a population transfer between the waves.  Our analysis enables us to interpret recent experimental observations of bright solitary matter waves.
\end{abstract}

\maketitle

Solitons are intriguing nonlinear wavepackets that propagate without dispersion due to the counteracting nonlinearity of the medium.  They occur widely in nature, for example, in nonlinear optics, water, mechanics,
biological systems, astrophysics, geology and particle physics
\cite{soliton_book}. In recent years, solitonic matter waves have been realised in atomic Bose-Einstein condensates (BECs) in bright \cite{khaykovich,strecker,cornish_new}, dark \cite{DS} and gap \cite{gap} forms.  Here the atomic interactions introduce a nonlinearity to the system such that, at zero-temperature and on a mean-field level, the BEC satisfies a cubic nonlinear Schr{\"o}dinger equation \cite{dalfovo} and supports the well-known one-dimensional soliton solutions derived by Zakharov and Shabat \cite{zakharov}.  Bright solitons are supported by attractive atomic interactions and manifest themselves as self-trapped droplets of matter.  Although these solitons are technically one-dimensional solutions, the analogous structure in 3D is a bright solitary wave (BSW) \cite{perez_garcia,salasnich,carr_castin,parker_NJP}.  Under radial confinement, e.g. in an atomic waveguide potential, these states are
self-trapped in the axial direction. Due to their novel properties
of self-trapping and shape-preservation, bright solitary
matter waves hold strong advantages for applications such as for
atom optics and interferometry \cite{strecker}. However, attractive
BECs in 2D/3D suffer from a collapse instability when the atom
number becomes too large
\cite{salasnich,carr_castin,parker_NJP,gammal}.

One of the most interesting aspects of solitons are their collisions and recent experiments have probed the collisions of matter wave BSWs 
\cite{strecker,cornish_new}. Cornish {\it et al.} 
\cite{cornish_new} generated multiple matter-wave BSWs \cite{creation} and observed their dynamics in a trap.  In particular, two BSWs were observed to oscillate in anti-phase along the axial direction of the harmonic trap for over $3$~s, colliding in the trap centre approximately 40 times.  The stability of these dynamics was somewhat surprising given the almost three-dimensional trap geometry and the combined population of the BSWs exceeding the critical atom number for collapse. In the 1D limit, soliton interactions
have been well-expounded, with the force between two solitons known to depend
sinusoidally on their relative phase $\Delta \phi$ \cite{gordon}. For 
$\Delta\phi=0$, the symmetric wavepackets can overlap freely,
leading to an `attractive' interaction, while for $\Delta \phi=\pi$
the asymmetric wavefunction prevents overlap and leads to a
`repulsive' interaction \cite{carr_PRE}.  In this 1D limit the collisions
are always elastic and the relative phase does not change the final
outgoing states \cite{gordon,carr_PRE}. In contrast colliding 
BSWs in 3D can form a high-density state that is unstable to collapse
\cite{salasnich,adhikari,carr,baizakov,khaykovich2}. Here, the relative
phase plays a crucial role, with an `attractive' $\Delta \phi=0$
collision being prone to the collapse instability and a `repulsive' $\Delta
\phi=\pi$ collision predicted to negate collapse effects
\cite{salasnich,carr}. Recent work \cite{khaykovich2} also indicates
that the collisional speed is also crucial in the stability of BSW
collisions.  In the majority of theoretical studies of BSW collisions,  approximations have been employed to simplify the approach, for example, by reduction of the 3D dynamics to an effective 1D model \cite{salasnich,carr,khaykovich2} or the use of a variational approach \cite{khawaja}.  Although full 3D simulations of BSW collisions have been made \cite{adhikari,baizakov}, a detailed study of the relevant parameter space is still lacking.  We note that analogous effects are observed for optical solitons in saturable nonlinear media \cite{optics} and solitonic Q-balls in particle physics \cite{particle}, including soliton fusion and annihilation.

In this work we theoretically analyse the rich behaviour of the
collisions of three-dimensional bright solitary matter waves. This is performed through extensive numerical simulations of the 3D Gross-Pitaevskii equation.  We
elucidate how the collisions depend on the key
parameters, namely the relative phase, interaction strength and
timescale of the collision. We apply our analysis to the 
recent experiment of 
Cornish {\it et al.} \cite{cornish_new} (henceforth referred to as the {\em JILA experiment}) and 
give strong evidence to the existence of a 
$\pi$-phase difference between the experimental BSWs.  

In the limit of ultra-cold temperature the mean-field `wavefunction'
of the BEC $\psi({\bf r},t)$ is well-described by the
Gross-Pitaevskii equation (GPE) \cite{dalfovo},
\begin{equation}
i\hbar \frac{\partial \psi}{\partial
t}=\left[-\frac{\hbar^2}{2m}\nabla^2 + \frac{m}{2}\omega_r ^2
\left(r^2 + \lambda^2 z^2\right)+\frac{4\pi\hbar^2a_{\rm
s}}{m}|\psi|^2 \right]\psi,
\end{equation}
where $m$ is the atomic mass and $a_{\rm s}$ is the {\it s}-wave
scattering length ($a_{\rm s}<0$ for the case of
attractively-interacting BECs considered here). The confining
potential is cylindrically-symmetric and harmonic, with radial
frequency $\omega_r$ and the axial frequency defined via the trap
ratio $\lambda=\omega_z/\omega_r$. The mean-field wavefunction
satisfies $\psi({\bf r},t)=\sqrt{n({\bf r},t)}\exp(i\theta({\bf
r},t))$, where $n({\bf r},t)$ is the atomic density and $\theta({\bf
r},t)$ is the condensate phase. The GPE provides an excellent model
of mean-field effects in BECs and has accurately predicted the {\em
onset} of collapse of an attractive BEC \cite{parker_NJP,gammal}.
However, the basic GPE is insufficient to model post-collapse
dynamics where higher-order effects, such as three-body losses,
become considerable and more sophisticated models must be employed, e.g. \cite{wuster}. We simulate the BSW dynamics by numerical propagation of 
Eq.~(1) on a cylindrically-symmetric spatial grid using 
the Crank-Nicholson propagation technique \cite{minguzzi}.

It is convenient to define a dimensionless
interaction parameter $k=N|a_{\rm s}|/a_r$, where
$a_r=\sqrt{\hbar/m\omega_r}$ is the radial harmonic oscillator
length.  When $k$ exceeds a critical value $k_{\rm c}$ the system is
unstable to collapse \cite{salasnich,carr_castin,parker_NJP}.  For a BSW ($\lambda=0$), it has been predicted that $k_{\rm c}\approx 0.67$ \cite{parker_NJP}, while the presence of an axial confining potential ($\lambda>0$) weakly reduces $k_c$, 
e.g., $k_{\rm c} \approx 0.64$ for $\lambda=0.4$ \cite{parker_NJP,gammal}.
Although our results are generic, they are presented in terms of the
parameters of the $^{85}{\rm Rb}$ JILA experiment \cite{cornish_new}. This
featured full 3D confinement defined by $\omega_r/2\pi=17.3$~Hz and
$\lambda=0.4$. Specifically, for $a_{\rm s}=-0.6~$nm, two BSWs were observed, with
a total measured atom number of $N=4000$. Allowing for approximately $500$ 
thermal atoms in the experimental measurement \cite{cornish_new}, we will assume 
each BSW to contain $N=1750$, giving $k=0.4$.

We first consider the simplest geometry of an axially-homogeneous
waveguide ($\lambda=0$) with finite radial trapping ($\omega_r>0$).  The BSW ground state has the approximate form $\psi(r,z)=\sqrt{N/2\pi \xi a_r^2}{\rm sech}(z/\xi)\exp(-r^2/2 a_r^2)$ \cite{salasnich,carr_castin,parker_NJP}. Here $\xi=1/\sqrt{4\pi n_0 |a_{\rm s}|}$ characterises the axial size of the BSW, where $n_0$ is the peak density.  We obtain the exact BSW ground state by numerical propagation of Eq.~(1) in imaginary
time \cite{minguzzi}. Our initial state consists of two
such solutions, well-separated at positions $z=\pm z_0$. Each
BSW is given a velocity kick $v_{\rm i}$ towards the origin (via
$\psi(z,r)\rightarrow \psi(z,r)\exp(imv_{\rm i} |z|/\hbar)$).
Furthermore, a phase difference $\Delta \phi$ is imprinted between
the BSWs.  

\begin{figure}
\centering
\includegraphics[width=11.cm,clip=true]{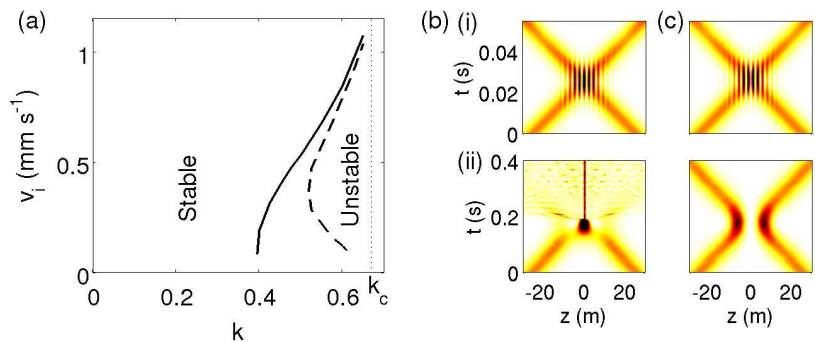}
\caption{(a) Stability of BSW collisions in an axially homogeneous waveguide for $\Delta \phi=0$ (solid line) and
$\pi$ (dashed line) as a function of $k=N|a_{\rm
s}|/a_r$ and speed $v_{\rm i}$. To the left (right) of the lines, the collisions are stable (unstable). The dotted line indicates $k_{\rm c}$ for an isolated BSW. (b)  Density plots show the evolution of a BSW ($k=0.4$) collision with $\Delta \phi=0$ at (i) high speed $v_{\rm i}=1~{\rm mm~s}^{-1}$ and (ii) low speed $v_{\rm i}=0.1~{\rm mm~s}^{-1}$.  Here we plot the axial density, integrated over $r$. (c)  Same as (b) but for $\Delta \phi=\pi$.} \label{fig1a}
\end{figure}

Just as the interaction parameter $k$ determines the stability of an
isolated BSW it is a crucial factor in the stability of their
collisions. Salasnich {\it et al.} predict that a $\Delta \phi=0$
collision is unstable for $k \geq 0.472$, based on the nonpolynomial
GPE and a BSW ansatz \cite{salasnich}.  Moreover, it is
predicted that a $\Delta \phi=\pi$ collision is always stable to collapse
\cite{salasnich,carr,khawaja}. We have performed extensive numerical
simulations of BSW collisions to map out the parameter space of $k$
and $v_{\rm i}$ for phase differences of $\Delta \phi=0$ and $\pi$.  The results are presented in Fig. \ref{fig1a}(a).  The solid and dashed lines mark the transition between stable and unstable collisions for $\Delta \phi=0$ and $\Delta \phi=\pi$, respectively. During an unstable collision the peak density increases above the threshold
for collapse and triggers a collapse
instability which destroys the BSWs. An example of an unstable
collision is shown in Fig.~\ref{fig1a}(b)(ii), for $\Delta \phi=0$. In
the stable regime, the collisions are elastic, with the BSWs emerging with the
same speed and shape as the incoming BSWs. Typical stable collisions
are shown in Figs. \ref{fig1a}(b)(i), (c)(i) and (c)(ii).

In general the collisions are stable for low values of $k$ but
become unstable as $k$ increases towards $k_{\rm c}$. At low speeds
the extent of the unstable region is dependent on the
relative phase $\Delta \phi$, as illustrated by comparing Figs. \ref{fig1a}(b)(ii) and (c)(ii), with
$\Delta \phi=\pi$ collisions being  more stable since the overlap of
the BSWs is prevented. However, at large impact speeds the stability
of the collisions is independent of $\Delta \phi$, as demonstrated
by the similarity between Figs. \ref{fig1a}(b)(i) and (c)(i).  Note that the number of collisional fringes increases with speed \cite{salasnich} and is always even (odd) for $\Delta \phi=0$ ($\pi$).  For $\Delta \phi=0$
and in the limit $v_{\rm i} \rightarrow 0$, the
collisions become unstable for $k\gtsimeq 0.4$.
As $v_{\rm i}$ is increased, this threshold shifts monotonically to
higher values of $k$, and for $v_{\rm i} \sim 1~{\rm mm~s}^{-1}$ it
is close to $k_{\rm c}$.  For $\Delta \phi=\pi$ and in the limit $v_{\rm i} \rightarrow 0$, the
collisions become unstable for $k \gtsimeq 0.6$. That is, at low
speeds, a $\pi$-phase difference collision can support a much greater interaction strength, and therefore number of atoms, than a $0$-phase difference collision. 

\begin{figure}
\centering
\includegraphics[width=11.cm,clip=true]{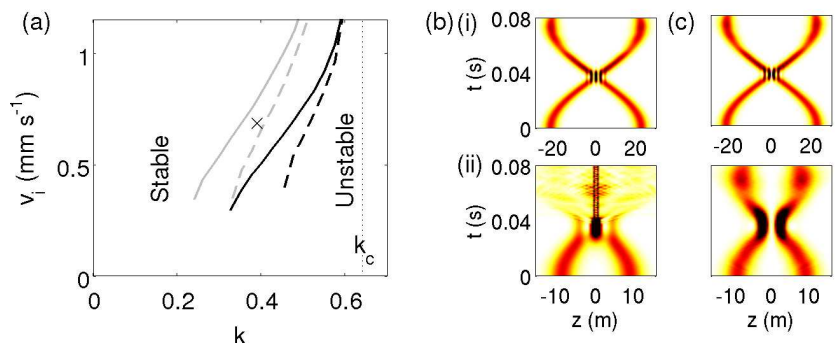}
\caption{(a) Stability of BSW collisions in a $\lambda=0.4$ trap for (i) $\Delta \phi=0$ and
(ii) $\pi$, as a function of $k$ and $v_{\rm i}$. The lines represent the transition between stable and unstable collisions.  Solid (dashed) lines correspond to $\Delta \phi=0$ ($\Delta \phi=\pi$) collisions, and bold (grey) lines represent the collisional stability after one ($40$)  collisions. We cannot present
results for low $v_{\rm i}$ since the wavepackets overlap initially.
The position of the JILA experiment is shown by the cross and the dotted line indicates $k_{\rm c}$. (b)-(c)
Density plot of a single ($k=0.4$) collision with (b) $\Delta \phi=0$ and (c) $\Delta
\phi=\pi$ for initial positions (i) $z_0=23~\mu$m ($v_{\rm i}\approx 1~{\rm mm~s^{-1}}$) and (ii) $z_0=9~\mu$m ($v_{\rm i}\approx 0.4~{\rm mm~s^{-1}}$). } \label{fig1b}
\end{figure}

Although a $\pi$-phase difference suppresses collapse between the colliding BSWs, it
does not completely prevent it. The density profile of the BSWs
alters as they approach, leading to an enhanced peak density. As
$v_{\rm i}$ is increased this enhancement becomes larger and thus
the threshold for instability initially moves to lower values of
$k$. However, for $v_{\rm i} \gtsimeq 0.4 {\rm mm~s}^{-1}$, the
threshold moves to higher values of $k$ and ultimately approaches
$k_{\rm c}$.

At the transition between stable and unstable collisions, we observe a
narrow region of inelastic collisions, where the outgoing BSWs have
modified shape and speed due to the occurrence of partial collapse during the
collision.  This typically manifests in the excitation of collective modes in the outgoing BSWs.

We now consider the presence of axial trapping \footnote{Although these states are not strictly solitonic when under external axial trapping, we will continue to refer to them as BSWs.}.  This is known to lower the critical point for collapse of the ground state \cite{parker_NJP,gammal}. Our initial state consists of two ground
state wavepackets, positioned off-centre in the trap at $z=\pm
z_0$ and given a phase difference $\Delta \phi$.  Density plots showing the typical evolution of a BSW collision in a trap are shown in Fig.~\ref{fig1b}(b) and (c). The BSWs accelerate down the trap and collide at the origin with approximate speed $v_{\rm i}=\lambda \omega_r z_0$.    First we consider the stability after just one collision in the trap.  We observe qualitatively similar features to the $\lambda=0$ case.  For low speed and $\Delta \phi=0$, collapse can occur (Fig.~\ref{fig1b}(b)(ii)) while a $\pi$-phase difference can prevent collapse  (Fig.~\ref{fig1b}(c)(ii)).  At higher speed, the outcome becomes independent of $\Delta\phi$, as illustrated in Figs.~\ref{fig1b}(b)(i) and (c)(i).  This insensitivity to the relative phase at high impact speeds allows one to adopt a simple particle model to describe this regime \cite{martin}.
Figure~\ref{fig1b}(a) presents the relevant parameter space for $\Delta\phi=0$ (bold solid line) and $\Delta \phi=\pi$ (bold dashed line) collisions.
Comparison to Fig.~\ref{fig1a}(a) shows that the axial
trapping shifts the threshold for instability to lower values of
$k$.  Note that for more (less) spherical traps the transition lines get shifted to lower (higher) values of $k$.  Since multiple collisions can occur in the trap, we also probe the collisional stability after $40$ collisions (grey  lines in Fig.~\ref{fig1b}(a)) for $\Delta \phi=0$ (solid line) and $\pi$ (dashed line). The transition between stable/unstable collisions gets shifted to even lower values of $k$.   

In the JILA experiment, the two BSWs can be clearly resolved when they are at the outer turning points of their oscillatory motion in the harmonic trap, revealing their maximum displacement to be $z_0 \approx 16~\mu$m.  We estimate their collisional speed to be $v_{\rm i}=\lambda \omega_r z_0=0.7~{\rm mm~s}^{-1}$.  The point corresponding to the JILA experiment is indicated in Fig.~\ref{fig1b}(a) by the cross.  For $\Delta \phi=0$ we see that the JILA BSWs are stable after one collision but are destroyed by $40$ collisions.  In contrast for $\Delta \phi=\pi$ the JILA BSWs are stable even after $40$ collisions.  Given that over $40$ stable oscillations were observed in the JILA experiment, this gives strong evidence towards the existence of a $\pi$-phase difference in the experiment.

We propose that the stability of BSW collisions is a play off
between the timescale over which the BSW interact $t_{\rm int}$ and the characteristic time for collapse to occur $t_{\rm col}$: if $t_{\rm int} < t_{\rm col}$ then the passing
wavepackets do not have time to
collapse; however, if $t_{\rm int} > t_{\rm col}$, the collapse has
sufficient time to develop during the collision. Based on this idea, the onset of instability in high speed collisions will be dominated by the collapse timescale $t_{\rm col}$ and this would imply that onset of instability is independent of $\Delta \phi$, as observed in our simulations.  Consider just one $\Delta \phi=0$ collision in the JILA system ($k=0.4$, $\lambda=0.4$).  From Fig.~\ref{fig1b}(a), the critical speed for collapse is $v_{\rm c} \approx 0.6 ~{\rm mm~s}^{-1}$.  We can estimate the interaction time at this speed as $t_{\rm int}=\xi/v_{\rm c}$. From \cite{parker_NJP} the size of the JILA wavepacket is $\xi\approx 3~\mu$m, giving $t_{\rm int}\approx 5~$ms.  The characteristic timescale for interaction-induced collapse has been
measured experimentally in this system to be $5(1)~$ms \cite{donley}.  In other words, at the point at which we observe the onset of unstable collisions, $t_{\rm int}\approx t_{\rm col}$.  This gives evidence to support this proposal.  Due to the lack of experimental data and an accurate theoretical model for collapse times \cite{wuster}, we cannot currently extend upon this prediction.

\begin{figure}[t]
\centering
\includegraphics[width=8cm,clip=true]{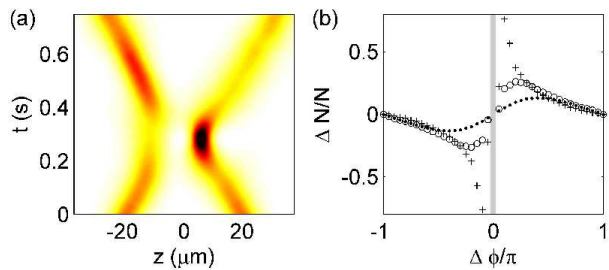}
\caption{(a) Density plot of a collision of two ($k=0.4$) BSWs for $v_{\rm i}=0.05~{\rm mm~
s}^{-1}$ and $\Delta \phi=0.5\pi$. (b) Population transfer ($\Delta
N/N$) versus $\Delta \phi$ for speeds of $v_{\rm i}=0.025$ (crosses),
$0.05$ (circles) and $0.1~{\rm mm~s}^{-1}$ (dots) for $k=0.4$. For
$\Delta \phi \approx 0$ (shaded region) the collisions are unstable
to collapse due to maximal overlap of the waves.} \label{fig3}
\end{figure}

For the cases of $\Delta \phi=0$ and $\pi$ considered so far, the
evolution of the density is symmetric about the origin throughout the dynamics. For
intermediate phases $0<\Delta\phi<\pi$, the collisional density
becomes asymmetric \cite{gordon,carr_PRE}. Using an effective 1D
GPE, valid under strong quasi-1D confinement, Khaykovich
{\it et al.} \cite{khaykovich2} have indicated the transfer of atoms
between the colliding waves.  Here we will consider the full 3D dynamics of BSW collisions
in a $\lambda=0$ system for the whole range of phase differences
$-\pi\leq\Delta\phi\leq\pi$. An example is shown in
Fig.~\ref{fig3}(a) for two identical BSWs ($k=0.4$) colliding
with relative phase $\Delta \phi=0.5\pi$. The asymmetric collision
induces a considerable population transfer between the waves,
generating a highly-populated and lowly-populated BSW. To
conserve momentum, the highly (lowly)-populated BSW travels at
reduced (increased) speed.

\begin{figure}[t]
\includegraphics[width=12cm,clip=true]{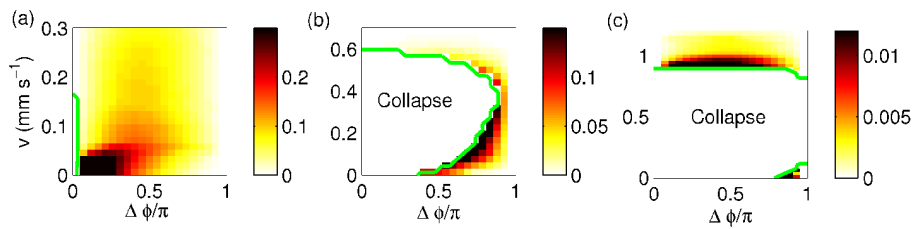}
\centering
\caption{Population transfer ($\Delta N/N$) between two colliding BSWs in an 
axially-homogeneous waveguide ($\lambda=0$) in the parameter space
of relative phase $\Delta \phi$ and speed $v_{\rm i}$ for (a) $k=0.4$, (b) 
$k=0.5$ and (c) $k=0.6$.} \label{fig4}
\end{figure}

We quantify the population transfer after the collision by the ratio
$\Delta N/N$, where $\Delta N$ is the number of atoms transferred
and $N$ is the initial population of each BSW.  In Fig.~\ref{fig3}(b)
we plot $\Delta N/N$ as a function of relative phase $\Delta \phi$
for the collision of two $k=0.4$ BSWs at various
speeds. At the highest speed (dots) the population transfer varies
sinusoidally with $\Delta \phi$, reaching a peak at $\Delta
\phi=\pi/2$. However, at lower speed (circles) the population
transfer becomes skewed, with the peak moving towards $\Delta
\phi=0$. At the lowest speed presented (crosses) the population
transfer appears to diverge as $\Delta \phi \rightarrow 0$.  This information is also shown in
Fig.~\ref{fig4}(a) which maps out the population transfer in the
parameter space of $v_{\rm i}$ and $\Delta \phi$ for $k=0.4$.  Here we clearly see that the population transfer is maximal for low speeds and decays with $v_{\rm i}$. 
Around $\Delta \phi =0$ and at low $v_{\rm i}$, we find a small
region where the BSW overlap is so great that a collapse instability
is induced. For an increased value of $k=0.5$ this unstable region
becomes larger, as shown in Fig.~\ref{fig4}(b), with collisions only
being stable for all phases when $v_i>0.6~{\rm mm~s}^{-1}$. The region of
collapse instability increases even further for $k=0.6$
[Fig.~\ref{fig4}(c)], where stable collisions for all $\Delta \phi$
occur only for $v_i>0.9$mms$^{-1}$.  As shown in Figs.~\ref{fig3} and \ref{fig4}, the magnitude of the population transfer can be a large fraction of the total population and is expected to be experimentally detectable.

We have performed an approximate two-mode analysis of this problem,
similar to that performed for BECs in static double well potentials
\cite{sakellari}, but with each mode being propagated at constant speed $v_{\rm i}$ through each other.  When closely positioned, Josephson-like tunnelling occurs between the states, with the final population transfer depending sinusoidally on $\Delta \phi$ and decaying exponentially with $v_{\rm i}$ due to the reduced interaction time.  This is in qualitative agreement with the GPE simulations, suggesting that Josephson-like tunnelling is the key process. However, the two-mode analysis over-estimates the amplitude of the population transfer and does not describe the observed divergent behaviour or the regions of collapse.  Since the two-mode analysis grossly fails to describe the collisional state of the BSWs (e.g. the formation of fringes), this is not surprising.

In summary, we have shown that the collisions of bright solitary waves exhibit
rich and non-trivial behaviour, not present for 1D solitons. High-density collisions
can induce collapse, depending on the collision time $t_{\rm int}$
(and therefore the collisional speed) relative to the collapse time
$t_{\rm col}$. For $t_{\rm int} > t_{\rm col}$ the BSWs are
completely destroyed by a catastrophic collapse, with the presence of a $\pi$-phase difference between the waves
suppressing this instability.  For $t_{\rm int} < t_{\rm col}$, the
collisions are elastic and independent of relative phase.
Using our analysis we show that the experimental observations of
long-lived `soliton' oscillations by Cornish {\it et al.}
\cite{cornish_new} require the existence of a $\pi$-phase difference.
Furthermore, we reveal a Josephson-like population transfer between the colliding BSWs that depends sensitively on relative phase and can be a large proportion of the total number.  As such, this effect may provide a route to matter-wave interferometry with solitons, which will be examined in future work.

\ack We acknowledge the UK EPSRC (NGP/SLC/CSA), Royal Society (SLC),
University of Melbourne (NGP/AMM) and ARC (NGP/AMM) for support. We
thank S. A. Gardiner, J. Brand and D. H. J. O'Dell for discussions.

\end{document}